\documentclass[twoside,11pt,preprint]{article}

\usepackage{jmlr2e}
\usepackage{pythonhighlight}
\usepackage{amsmath}
\usepackage{tabularx}
\usepackage{makecell}
\usepackage{geometry}
\usepackage{graphicx}
\usepackage{amsmath, amssymb}
\graphicspath{ {figures/} }

\usepackage{float}
\usepackage{cleveref}

\floatstyle{plain}
\newfloat{pyprogram}{thp}{lop}
\floatname{pyprogram}{Example Code}
\crefname{pyprogram}{Example Code}{Example Code} 

\usepackage{lastpage}
\jmlrheading{24}{2024}{1-\pageref{LastPage}}{04/23; Revised 5/22}{9/22}{21-0000}{Sankalp Gilda}

\ShortHeadings{tsbootstrap}{Gilda}
\firstpageno{1}

\begin{document}

\title{\texttt{tsbootstrap}: Enhancing Time Series Analysis with Advanced Bootstrapping Techniques}

\author{\name Sankalp Gilda \email sankalp.gilda@gmail.com \\
       \addr ML Collective\\
       San Francisco, CA 98195-4322, USA
       \AND
       \name Benedikt Heidrich \email benedikt.heidrich@sktime.net \\
       \addr sktime project
       \AND
       \name Franz Kiraly \email franz.kiraly@sktime.net \\
       \addr sktime project}
\editor{My editor}

\maketitle

\begin{abstract}
In time series analysis, traditional bootstrapping methods often fall short due to their assumption of data independence, a condition rarely met in time-dependent data. This paper introduces \texttt{tsbootstrap}, a \texttt{python} package designed specifically to address this challenge. It offers a comprehensive suite of bootstrapping techniques, including Block, Residual, and advanced methods like Markov and Sieve Bootstraps, each tailored to respect the temporal dependencies in time series data. This framework not only enhances the accuracy of uncertainty estimation in time series analysis but also integrates seamlessly with the existing \texttt{python} data science ecosystem, making it an invaluable asset for researchers and practitioners in various fields.
\end{abstract}


\section{Introduction}\label{sec:introduction}
Time series data, characterized by their sequential observations, are the backbone of critical decision-making in various scientific and economic fields. Unlike static or cross-sectional datasets, time series data are dynamic, evolving over time and exhibiting intricate dependencies that reflect the complexities of the real world. This dynamic nature poses unique challenges, particularly when it comes to predicting future events (forecasting) and assessing the reliability of these predictions (uncertainty quantification). Traditional statistical methods, designed for independent and identically distributed (IID) data, falter in this context, as they overlook the temporal correlations that are fundamental to time series data. This gap in the statistical toolkit highlights a pressing need for methodologies that can adeptly navigate the intricacies of time series, preserving their inherent structure while facilitating rigorous analysis and inference.

Bootstrapping \citep{bootstrap_jackknife_efron, practical_guide_bootstrap} offers a powerful tool for estimating the variability of a statistic without making stringent assumptions about the underlying data distribution. However, the classic IID bootstrapping approach, which resamples data points as if they were independent, is ill-suited for time series where each observation is a chapter in an ongoing story, not an isolated event. To capture the essence of time series, bootstrapping must adapt, preserving the internal structure and dependencies within the data.

Addressing this need, various time series bootstrapping methods have emerged, each designed to respect the chronological integrity of the data while enabling robust uncertainty quantification \citep{kreiss_lahiri_time_series_bootstrap}. Yet, the software implementation of these methods presents its own set of challenges. They require not just statistical acumen but also computational finesse, demanding algorithms that can handle the complexity of time series data while being accessible to the broader scientific and industrial community.

\texttt{tsbootstrap} \citep{tsbootstrap_zenodo} emerges as a response to the complexities and specific needs in time series analysis. It is a \texttt{python} package designed to provide a range of bootstrapping methods tailored for time series data. This package includes traditional techniques such as Block Bootstrap \citep{block_bootstrap_orig} and extends to more sophisticated methods like Markov and Sieve Bootstrap \citep{sieve_bootstrap}. Each method within \texttt{tsbootstrap} is implemented to respect and preserve the chronological order and inherent correlations of time series data, ensuring more accurate and reliable bootstrapping.

We have designed \texttt{tsbootstrap} as a versatile and user-friendly tool to empower both academic researchers and industry practitioners, offering them new capabilities to navigate and interpret the complexities of time series data with unprecedented precision and insight. Specifically, we fill in two existing gaps:
\begin{itemize}
    \item \textit{Framework for Enhanced Uncertainty Quantification:} \texttt{tsbootstrap} provides a robust framework that significantly advances the practice of uncertainty quantification in time series analysis. This is particularly crucial in fields such as finance, where accurate risk assessment is essential; in meteorology, where predictive accuracy can have wide-ranging implications; and in epidemiology, where understanding uncertainty can inform public health decisions.

    \item \textit{Seamless Integration with the Time Series Analysis Ecosystem:} The package is meticulously designed to complement and enhance the existing \texttt{python} data science ecosystem. By ``time series analysis ecosystem'', we refer to the collection of tools, libraries, and methodologies that are commonly employed in the analysis of time series data. This includes integration with popular libraries like \texttt{sktime} \citep{sktime_zenodo} and \texttt{pandas} \citep{pandas_zenodo}, ensuring that \texttt{tsbootstrap} is not only a standalone tool but also a synergistic component that enhances the capabilities of other time series analysis tools.

\end{itemize}

In the rest of this paper, we delve into \texttt{tsbootstrap} in more depth, offering a comprehensive examination of its theoretical underpinnings, implementation details, and the breadth of its application. We demonstrate its utility in elevating time series analysis through case studies and comparative analyses, emphasizing its role as an essential instrument for researchers and practitioners. In subsequent sections we provide a structured exploration. In Section \ref{sec:bootstrapping} we introduce bootstrapping and its significance. In Section \ref{sec:tsbootstrap}, we focuse on \texttt{tsbootstrap}'s design principles and functionalities. In Section \ref{sec:supported bootstrap algorithms} we detail the supported bootstrap algorithms. In Section \ref{sec:integration_with_sktime}, we illustrate practical usage, integration with \texttt{sktime}, and extension capabilities. Finally, we conclude in Section \ref{sec:conclusion}, and outline the future trajectory of \texttt{tsbootstrap} in Section \ref{sec:future_work}.

\begin{table}
\centering
\begin{tabularx}{\textwidth}{| l | X | X |}
\hline
Bootstrapping Method & Class (...\texttt{Bootstrap}) & Salient Features \\
\hline
Moving Block & \texttt{MovingBlock} & Suitable for series with significant intra-block dependencies; avoid for weak dependencies \\
\hline
Circular Block & \texttt{CircularBlock} & Ideal for seasonal/cyclical data; not recommended for non-cyclical data \\
\hline
Stationary & \texttt{StationaryBlock} & Best for data with varying dependency lengths; less effective for uniform dependency lengths \\
\hline
Non-Overlapping Block & \texttt{NonOverlappingBlock} & Effective for data with distinct segments; not suitable for continuous or highly dependent data \\
\hline
Tapered Block & \texttt{Bartletts}, \texttt{Hamming}, \texttt{Blackman}, \texttt{Tukey} & Mitigates edge effects using window functions; requires matching window functions to data characteristics \\
\hline
Residual & \texttt{WholeResidual}, \texttt{BlockResidual} & Evaluates model-based uncertainty; not intended for non-model analyses \\
\hline
Statistic-Preserving & \texttt{WholeStatisticPreserving}, \texttt{BlockStatisticPreserving} & Maintains key statistical properties; not for general purposes without specific preservation needs \\
\hline
Distribution & \texttt{WholeDistribution}, \texttt{BlockDistribution} & Leverages known distributions; inappropriate for unknown distributions \\
\hline
Markov & \texttt{WholeMarkov}, \texttt{BlockMarkov} & Suitable for Markovian time series; avoid for non-Markovian series \\
\hline
Sieve & \texttt{WholeSieve}, \texttt{BlockSieve} & Designed for autoregressive models; not recommended for non-autoregressive series \\
\hline
\end{tabularx}
\caption{Overview of bootstrapping classes in \texttt{tsbootstrap}.}
\label{tab:bootstrap_methods}
\end{table}

\section{Bootstrapping}\label{sec:bootstrapping}
Bootstrapping is a resampling technique fundamental to statistics, providing a way to estimate the distribution of a sample statistic (like the mean or variance) by resampling with replacement from the data. It allows statisticians to assess the reliability of inferential statistics, such as confidence intervals and significance tests, without relying on stringent assumptions about the underlying population distribution.

At its core, bootstrapping involves repeatedly drawing samples, typically thousands of times, from a single observed dataset. Each sample is drawn with replacement, meaning the same data point can appear multiple times in a bootstrap sample. By applying the statistical measure of interest to each resample, a distribution of these statistics is generated, offering insight into the variability and bias of the estimate.

To underscore the importance of bootstrapping in various statistical processes, consider the following key aspects:
\begin{itemize}
    \item \textit{Model-Free Inference:} Bootstrapping does not assume a specific parametric form for the data distribution, making it a non-parametric approach. This flexibility is crucial when the underlying distribution is unknown or complex, allowing statisticians to avoid potentially incorrect model assumptions.

    \item \textit{Assessment of Uncertainty:} It provides a straightforward method to estimate the confidence intervals and standard errors of estimates, which are vital for understanding the precision of statistical inferences.

    \item \textit{Validation of Models:} In model building, bootstrapping can be used to validate the stability and reliability of predictive models. It helps in assessing the model's variance and bias, offering insights into its generalizability.

    \item \textit{Complex Estimators:} Bootstrapping is particularly valuable for complex estimators where the theoretical distribution is difficult or impossible to derive. It allows for the empirical examination of the estimator's distribution.
\end{itemize}

Transitioning to the specific challenges of time series data, it's crucial to recognize that the inherent dependencies within these data sets necessitate a more nuanced approach to bootstrapping. Traditional IID bootstrapping techniques, which treat each data point as independent, are not suitable for time series data due to their sequential nature \citep{resampling_lahiri}. This misalignment highlights the need for specialized time series bootstrapping methods that respect the data's temporal structure, ensuring the integrity of the analysis.

\section{\texttt{tsbootstrap}}\label{sec:tsbootstrap}

\begin{figure}
    \centering
    \includegraphics[width=\textwidth]{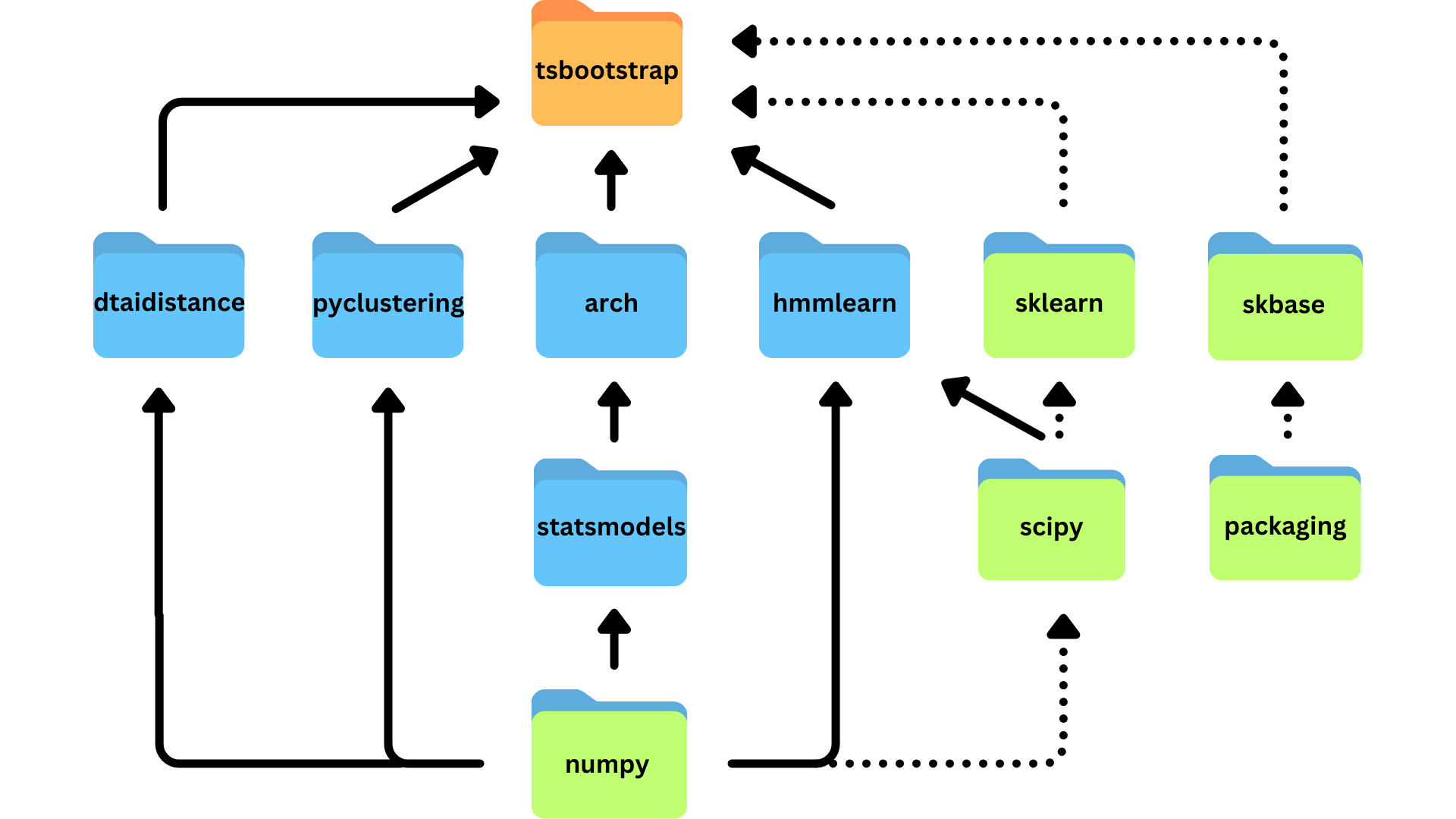}
    \caption{Optional (soft) and required (hard) dependencies for \texttt{tsbootstrap} v0.1.0. The latter are indicated via lime-colored folders and dashed arrows.}
    \label{fig:dependency_structure}
\end{figure}

\subsection{Design Principles}
\begin{enumerate}
    \item \textbf{Modularity and Extensibility}: The package's architecture is highly modular, allowing users to easily incorporate custom bootstrapping algorithms. This flexibility makes it an adaptable tool for a wide range of time series data applications. \texttt{tsbootstrap}'s design embraces the strategy pattern and composition, ensuring that new bootstrapping strategies can be easily added, promoting a collaborative environment for the community to contribute and extend the library.

    \item \textbf{User-Centric API}: The API is crafted with user experience in mind, ensuring simplicity in performing complex bootstrapping tasks. This design principle is evident in the package's comprehensive documentation and illustrative examples provided on the GitHub repository. Following \texttt{scikit-learn}-like conventions, \texttt{tsbootstrap} provides a consistent API 
    , making it accessible and user-friendly for those accustomed to \texttt{scikit-learn}'s patterns.

    \item \textbf{Integration with the \texttt{python} Ecosystem}: \texttt{tsbootstrap} is built to seamlessly integrate with popular \texttt{python} libraries like \texttt{numpy} and \texttt{pandas}, enhancing its utility in the prevalent data science ecosystem. Moreover, its design facilitates easy integration with other \texttt{python} libraries, particularly \texttt{sktime}, allowing \texttt{tsbootstrap} to be used within existing pipelines, enhancing the robustness and reliability of time series analysis.

    \item \textbf{Alignment with \texttt{scikit-learn}'s Cross-Validation}: Mirroring \texttt{scikit-learn}'s cross-validation class, \texttt{tsbootstrap} makes a slight adjustment: it focuses on bootstrapping for time series data, replacing the \texttt{split} method with a \texttt{bootstrap} method. This approach is particularly suited for time series data, respecting and preserving their temporal structure during the resampling process, and providing more reliable estimates of model performance.
\end{enumerate}

\subsection{Core Functionalities}
\begin{enumerate}
    \item \textbf{Diverse Bootstrapping Methods}: We include various methods such as Block \citep{block_bootstrap_orig, stationary_bootstrap, tapered_block_bootstrap}, Residual, Statistic-Preserving, and Distribution Bootstrap. Each method is fine-tuned for specific types of time series data, offering users a tailored approach to their analytical needs.
    \item \textbf{Customization and Flexibility}: Users can customize various parameters like block size, sampling methods, and random seed settings. This level of customization is instrumental in conducting precise and controlled bootstrapping analyses.
    \item \textbf{Efficient Data Handling and Processing}: We employ advanced data handling techniques, ensuring efficient processing of large time series datasets, input data and parameter validation, and fine-grained logging.
\end{enumerate}

\subsection{Integration into the \texttt{python} ML Universe}

\texttt{tsbootstrap} is designed to integrate seamlessly with established and widely used libraries like \texttt{sklearn} \citep{sklearn, sklearn_zenodo} and \texttt{sktime} \citep{sktime_zenodo}.

\texttt{tsbootstrap} adopts the task-specific unified interface principle, and composable specification language. It also provides task independent interface points common to \texttt{sklearn} and \texttt{sktime}, such as \texttt{get\_params}, \texttt{set\_params}, \texttt{get\_fitted\_params}.

This high-level interoperability between \texttt{tsbootstrap}, \texttt{sklearn}, and adjacent framworks such as \texttt{sktime} is mediated by the \texttt{skbase} library \citep{skbase_zenodo}, providing shared interface points and base classes. This API integration also allows \texttt{tsbootstrap} to leverage \texttt{sktime}'s handling of time series specific issues such as forecasting and time-dependent splitting.

Interoperability is in-principle possible both ways, and methodologically advantageous: \texttt{sktime} can leverage \texttt{tsbootstrap} to make, or improve, probabilistic forecasts, via custom bootstrapping schemas. Conversely, \texttt{tsbootstrap} can use \texttt{sktime} forecasters or transformations to construct bootstrap schemas with specific model assumptions, e.g., forecast or model residual bootstraps.

Finally, this design of \texttt{tsbootstrap} emphasizes extensibility and community engagement, encouraging contributions that expand its range of bootstrapping techniques and enhance its integration with the broader \texttt{python} data science landscape. Using an extensible, \texttt{sklearn}-like interface mediated by \texttt{skbase}, users can easily create interface compatible bootstrap algorithms, to add to the core library, or in their own code bases. Details on extension are presented in Section \ref{sec:extension_template}.

\section{Supported Bootstrap Algorithms}\label{sec:supported bootstrap algorithms}
\texttt{tsbootstrap} offers an extensive array of bootstrapping techniques, specially designed for time series data to preserve their inherent temporal structures during the resampling process. Below we provide a detailed overview of the supported bootstrap algorithms within the package:

\begin{figure}
    \centering
    \includegraphics[width=.8\textwidth]{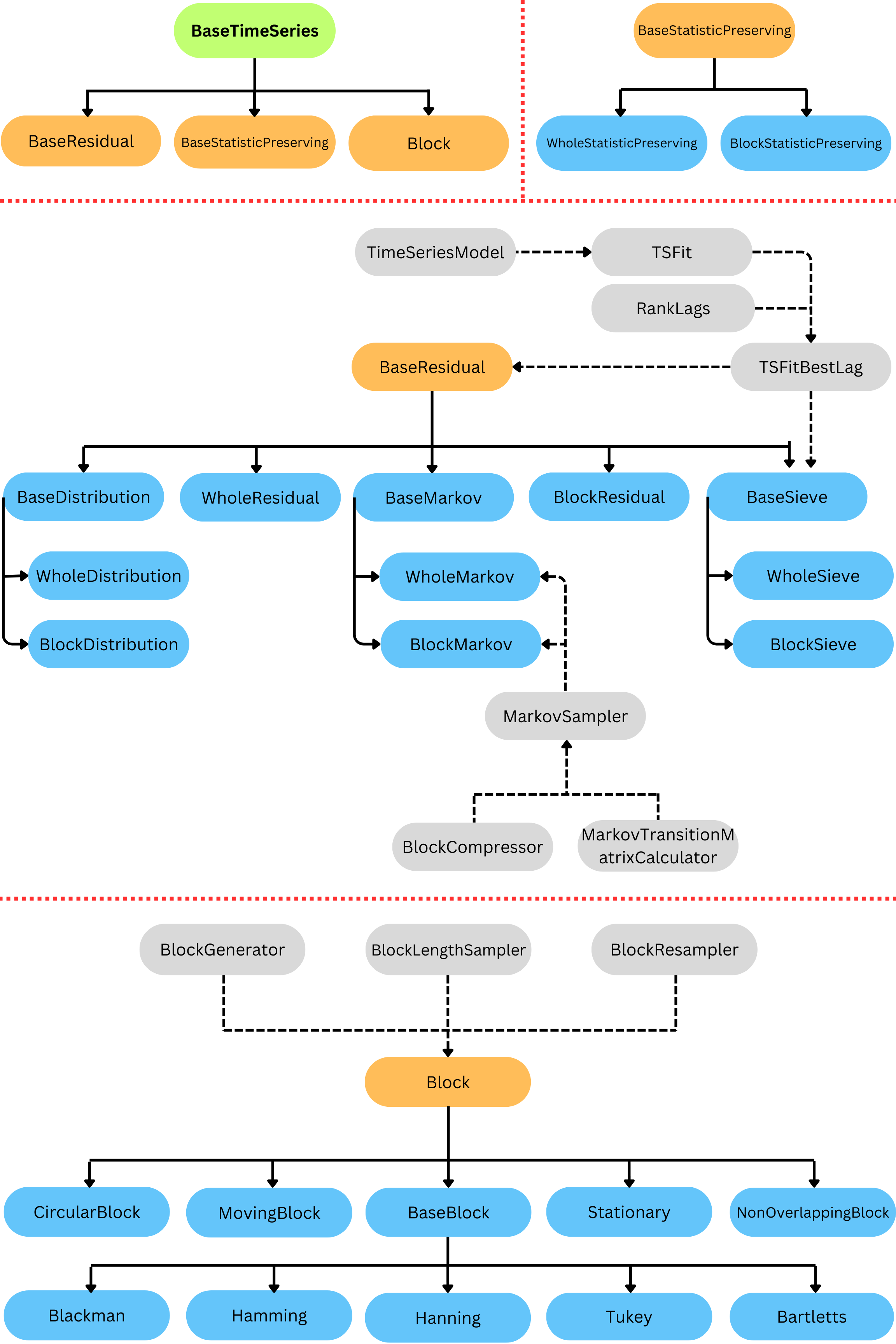}
    \caption{\textit{Top left:} \texttt{BaseTimeSeriesBootstrap} is the base-class, and has three main child classes, indicated in Orange. \textit{Top right, middle, and bottom:} Class structure for the three main child classes depicted from the top-left figure. {\color{gray}gray} blocks/classes are helper classes that are not exposed to the user. All classes/blocks in {\color{green}green}, {\color{orange}orange}, and {\color{blue}blue} have \texttt{Bootstrap} as the suffix in the package.}
    \label{fig:classes}
\end{figure}

\subsection{Block Bootstrap Methods}
For time series bootstrapping, Block bootstrap methods stand out as critical tools. These methods are indispensable because they allow the preservation of dependency structures within the data, a fundamental aspect that distinguishes time series from cross-sectional data. By resampling blocks of observations rather than individual data points, these methods maintain the chronological order and inherent dependencies, which are crucial for accurate time series analysis.

Here is a deeper look into the Block Bootstrap methods provided by \texttt{tsbootstrap}:

\begin{itemize}
    \item Moving Block Bootstrap (MBB): This method is particularly effective for time series with significant autocorrelation within contiguous segments. In MBB, overlapping blocks of data are resampled, ensuring preservation of the temporal dependence. This technique is especially useful in analyzing financial time series, where capturing the short-term dependencies is crucial for accurate forecasting and risk assessment.

    \item Circular Block Bootstrap (CBB): CBB extends the concept of MBB by treating the data as if it were circular, meaning that the end of the series connects back to the beginning. This method is particularly beneficial for handling seasonal or cyclical time series, as it allows the bootstrapping process to maintain the inherent cyclical properties, such as those found in meteorological or economic seasonal patterns.

    \item Stationary Block Bootstrap (SBB): This approach introduces a random element to the block length, providing a more dynamic resampling process. Unlike MBB or CBB, where the block size is fixed, Stationary Bootstrap \citep{stationary_bootstrap} randomly varies the length of each block according to a predetermined probability distribution. This method is adept at capturing both short- and long-term dependencies within the time series, offering a versatile approach that can adapt to various structures within the data.

    \item Non-Overlapping Block Bootstrap (NBB): NBB is a simpler variant where blocks of data are sampled in a non-overlapping manner. This method ensures that each data point is included only once in each bootstrap sample, reducing redundancy. NBB can be particularly effective for time series data with clear and distinct segments where the independence between blocks can be reasonably assumed.

    \item Tapered Block Bootstrap (TBB): TBB \citep{tapered_block_bootstrap} is an advanced variant that incorporates window functions to taper the blocks, reducing the potential discontinuities at the block edges. This method can be applied on top of the existing block bootstrap methods -- such as MBB, CBB, SB, and NBB -- as a base. The tapering process involves applying window functions like Hamming, Hanning, Blackman, or Tukey to the data blocks, which helps mitigate edge effects and improve the representativeness of the bootstrap sample. This technique is particularly useful in scenarios where the edge effects might introduce bias or when the time series exhibits strong seasonal components or trends. By employing different window functions, users can fine-tune the tapering to match the specific characteristics and requirements of their time series data, enhancing the flexibility and effectiveness of the bootstrapping process.
\end{itemize}

\subsection{Advanced Bootstrap Methods}
\texttt{tsbootstrap} includes a comprehensive suite of time series bootstrapping techniques, each designed for specific types of time series data:

\begin{itemize}
    
    \item \textit{Residual Bootstrap:} This method is particularly adept at handling model-based uncertainties \citep{kreiss_lahiri_time_series_bootstrap, residual_bootstrap}. After fitting a time series model to the data, the residuals (the differences between the observed values and the model's predictions) are bootstrapped. This approach allows for the assessment of the variability and uncertainty of the model predictions, accommodating a range of models from simple linear regressions to complex ARIMA models. The Residual Bootstrap is invaluable for validating the stability and accuracy of the predictions made by these time series models.

    \item \textit{Statistic-Preserving Bootstrap:} This technique ensures that the bootstrapped samples retain a user-defined key statistical property of the original dataset, be it mean, variance, or something else. It is particularly beneficial in applications where maintaining the original data's distributional characteristics is crucial for accurate analysis and inference. This method is essential for studies where the underlying distributional properties play a critical role in the analysis outcomes or model validations.

    \item \textit{Distribution Bootstrap:} When the underlying distribution of the time series is known or can be estimated, the Distribution Bootstrap method leverages this information to generate resampled data. This approach is ideal for datasets where specific distributional assumptions hold true, allowing for more precise and theoretically grounded inferences; in situations where the time series is believed to follow a known distributional pattern, aiding in more accurate modeling and forecasting.

    \item \textit{Markov Bootstrap:} Tailored for time series exhibiting Markovian properties, this method is suitable for data where the future state is dependent only on the current state and not on the sequence of events that preceded it. Commonly used in financial time series analysis, the Markov Bootstrap \citep{block_sieve_markov} can capture the stochastic processes underlying the financial markets, offering insights into the dynamics and potential future behaviors of these markets.

    \item \textit{Sieve Bootstrap:} This method is specifically designed for time series that can be modeled using autoregressive structures. By active on the coefficients of an autoregressive model, the Sieve Bootstrap provides a robust framework for analyzing and forecasting time-dependent data. This approach is particularly powerful in scenarios where the data exhibits autoregressive behavior, enabling analysts to make informed forecasts and understand the underlying dynamics of the time series \citep{sieve_bootstrap}.
\end{itemize}

Each of these methods demonstrates \texttt{tsbootstrap}'s capability to address the unique challenges presented by different types of time series data, making it a versatile tool in the field of time series analysis.

\subsection{Extension Template}\label{sec:extension_template}

\texttt{tsbootstrap} is meant to be extensible, designed to make it easy for power users to add new bootstrap algorithms, to the main code base, or (potentially closed or differently licensed) third party extensions.

For this, \texttt{tsbootstrap} follows the combined strategy/template design pattern also seen in \texttt{sktime} or \texttt{sklearn}, using the foundations of \texttt{skbase} for extension and testing:

\begin{itemize}
    
    \item \textit{Extension template:} implementers of new algorithms are provided with an extension template, i.e., a fill-in \texttt{python} file with step-by-step instructions on filling in a new algorithm.
    \item \textit{Extension contract:} boilerplate is abstracted away in the extension template - only a private \texttt{\_bootstrap} needs to be implemented, with simplified inputs, while class inheritance and boilerplate layering ensure public interface compliance.
    \item \textit{Contract checker:} extenders are provided with a compliance checker method, \texttt{check\_estimator}, which allows testing at runtime, and can be used in 3rd party testing setups for continuous API compliance checks.
\end{itemize}

Further details, including a 90 minute general-purpose tutorial to extending \texttt{skbase}-templated packages, can be found in the \texttt{skbase} documentation \citep{skbase_zenodo}.

\section{Usage and Integration with \texttt{sktime}}\label{sec:integration_with_sktime}

\subsection{Basic Usage}

In Figure \ref{fig:dependency_structure}, we detail the various building blocks and user-facing classes in \texttt{tsbootstrap}. Below, in \cref{code:tsbootstrap} and Figure \ref{fig:dependency_structure}, we demonstrate a simple usage of \texttt{MovingBlockBootstrap} using simulated data.

\begin{pyprogram}
    
\begin{python}[htbp]
from tsbootstrap import MovingBlockBootstrap
import numpy as np
import matplotlib.pyplot as plt

# Create custom time series data. While below is for univariate data,
# "tsbootstrap" can handle multivariate time series as well.
n_samples = 10
X = np.arange(n_samples)

# Instantiate the bootstrap object
n_bootstraps = 3
block_length = 3
rng = 42
mbb = MovingBlockBootstrap(
         n_bootstraps=n_bootstraps,
         rng=rng,
         block_length=block_length
      )

# Generate bootstrapped samples
return_indices = False
bootstrapped_samples = mbb.bootstrap(
    X, return_indices=return_indices)

# Collect bootstrap samples
X_bootstrapped = []
for data in bootstrapped_samples:
    X_bootstrapped.append(data)

X_bootstrapped = np.array(X_bootstrapped)

# Plot the bootstrapped samples and the original time series
plt.figure()
plt.plot(X, label="Original Time Series", color="black")
for i in range(n_bootstraps):
    _ = plt.plot(X_bootstrapped[i].reshape(-1,),
                 label=f"Bootstrapped Sample {i+1}",
                 ls="--")

plt.legend()
plt.xlabel("Time")
plt.ylabel("Value")
plt.show()

\end{python}
\caption{Example code snippet demonstrating the use of tsbootstrap} \label{code:tsbootstrap}

\end{pyprogram}

\begin{figure}[htbp]
    \centering
    \includegraphics[width=\textwidth]{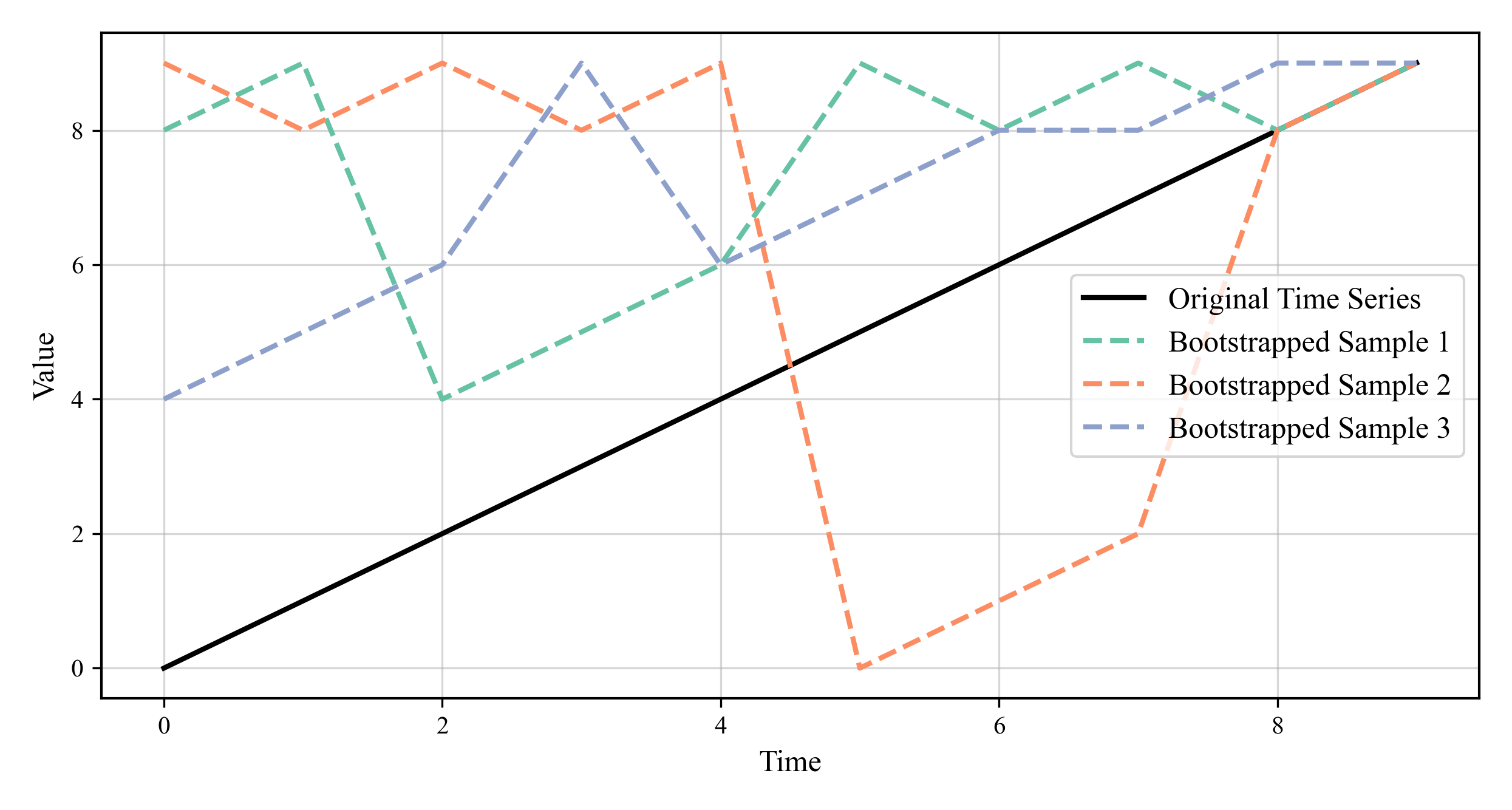}
    \caption{Output of \cref{code:tsbootstrap}.}
    \label{fig:tsbootstrap_example}
\end{figure}

\subsection{Integration with \texttt{sktime}}

\texttt{sktime} is a \texttt{python} library for time series providing unified interfaces for various tasks, such as classification, regression, or forecasting. For all of these tasks, bootrapping can improve the results. Thus, in the following, we first describe how \texttt{tsbootstrap} is integrated with \texttt{sktime}. We then show the combined usage of \texttt{sktime} and \texttt{tsbootstrap} with a probabilistic forecasting example. 

\subsubsection{\texttt{sktime}'s \texttt{TSBootstrapAdapter}}
\texttt{sktime}'s \texttt{TSBootstrapAdapter} follows the unified API design as the other transformers implemented in \texttt{sktime}. In other words, these methods provide \texttt{fit}, \texttt{transform}, \texttt{get\_params}, \texttt{set\_params} methods, and \texttt{\_\_init\_\_} as the constructor. Below we briefly present these methods.
\begin{description}
\item[\_\_init\_\_] is the constructor. It takes as parameters, the \texttt{tsbootstrap} object that should be used within \texttt{sktime}. 
\item[fit] fits the transformer. It is empty for this adapter.
\item[transform] creates the bootstraps. It takes as arguments, the time series that should be bootstrapped as $X$.
\item[get\_params] returns the current parameters of the adapter.
\item[set\_params] allows to set the different parameters.
\end{description}

This interface enables the usage of all \texttt{tsbootstrap} methods within \texttt{sktime} since it requires that an arbitrary \texttt{tsbootstrap} object is passed to the adapter and that all of these classes have the same interface. This interface consistency guarantees that the classes are exchangeable.  

\subsubsection{Usage: Probabilistic Forecasting}

Point forecasts comprises one value per time step. Thus, such forecasts do not quantify uncertainty. To quantify the uncertainty in forecasting, often probabilistic forecasts are generated. However, many forecasting algorithms do not support probabilistic forecasting directly. Thus, in \texttt{sktime}, there exist different methods to provide uncertainty quantification for point forecasts. One of these approaches that smoothly interacts with bootstrapping is the \texttt{BaggingForecaster}. 
The \texttt{BaggingForecaster} fits one model per bootstrap. During inference, each fitted model makes a forecast. Finally, out of this set of forecasts, the uncertainty can be derived. 

\Cref{code:sktime} shows an exemplary usage \texttt{tsbootstrap} and \texttt{sktime} for forecasting. First, the required modules are imported. Afterwards, the time series is loaded and split into train and test set. Third, the forecaster is initialised. In this example we use a \texttt{BaggingForecaster}. This forecaster takes as arguments a bootstrapper and a base forecaster. For the bootstrapper, we use \texttt{BlockResidualBootstrap} from \texttt{tsbootstrap}, which is wrapped using \texttt{TSBootstrapAdapter}, and as base forecaster, we use a SARIMA model. Fourth, we fit the forecaster using the training data. Finally, in the last step, we forecast the time series and also forecast intervals. 

\begin{pyprogram}
     \begin{python}
# Example code snippet demonstrating the use of tsbootstrap together with
# sktime
from sktime.forecasting.arima import ARIMA
from sktime.transformations.bootstrap import TSBootstrapAdapter
from sktime.forecasting.compose import BaggingForecaster
from sktime.transformations.series.detrend import Detrender, Deseasonalizer
from tsbootstrap import MovingBlockBootstrap
from sktime.utils.plotting import plot_series

from sktime.datasets import load_airline
from sktime.forecasting.model_selection import temporal_train_test_split

import matplotlib.pyplot as plt

# Load time series data with sktime or custom data with pandas
data = load_airline()
train, test = temporal_train_test_split(data, test_size=12)

# Initialize the bootstrapper with custom settings
adapter = TSBootstrapAdapter(
             BlockResidualBootstrap(
                MovingBlockBootstrap(20)
             )
          )

# Create a BaggingForecaster using the bootstrap adapter and a SARIMA model
forecaster = Deseasonalizer(sp=12, model="multiplicative") * Detrender() * 
             BaggingForecaster(adapter, 
                              ARIMA(order=(1, 1, 0), 
                              seasonal_order=(0, 1, 0, 12)),)

# Fit the forecaster
forecaster.fit(train)

# Perform predictions
pred = forecaster.predict(range(1, len(test) + 1))
pred_intervals = forecaster.predict_interval(range(1, len(test) + 1))

# Plot
plt.close("all")
fig, ax = plot_series(data, pred, pred_interval=pred_intervals, labels=["y", "y_pred"])
ax.set_xlabel("Time")
plt.show()

\end{python}

\caption{Exemplary usage of \texttt{tsbootstrap} together with \texttt{sktime} to perform a forecast on the airline dataset.} \label{code:sktime}

\end{pyprogram}

\begin{figure}[htbp]
    \centering
    \includegraphics[width=\textwidth]{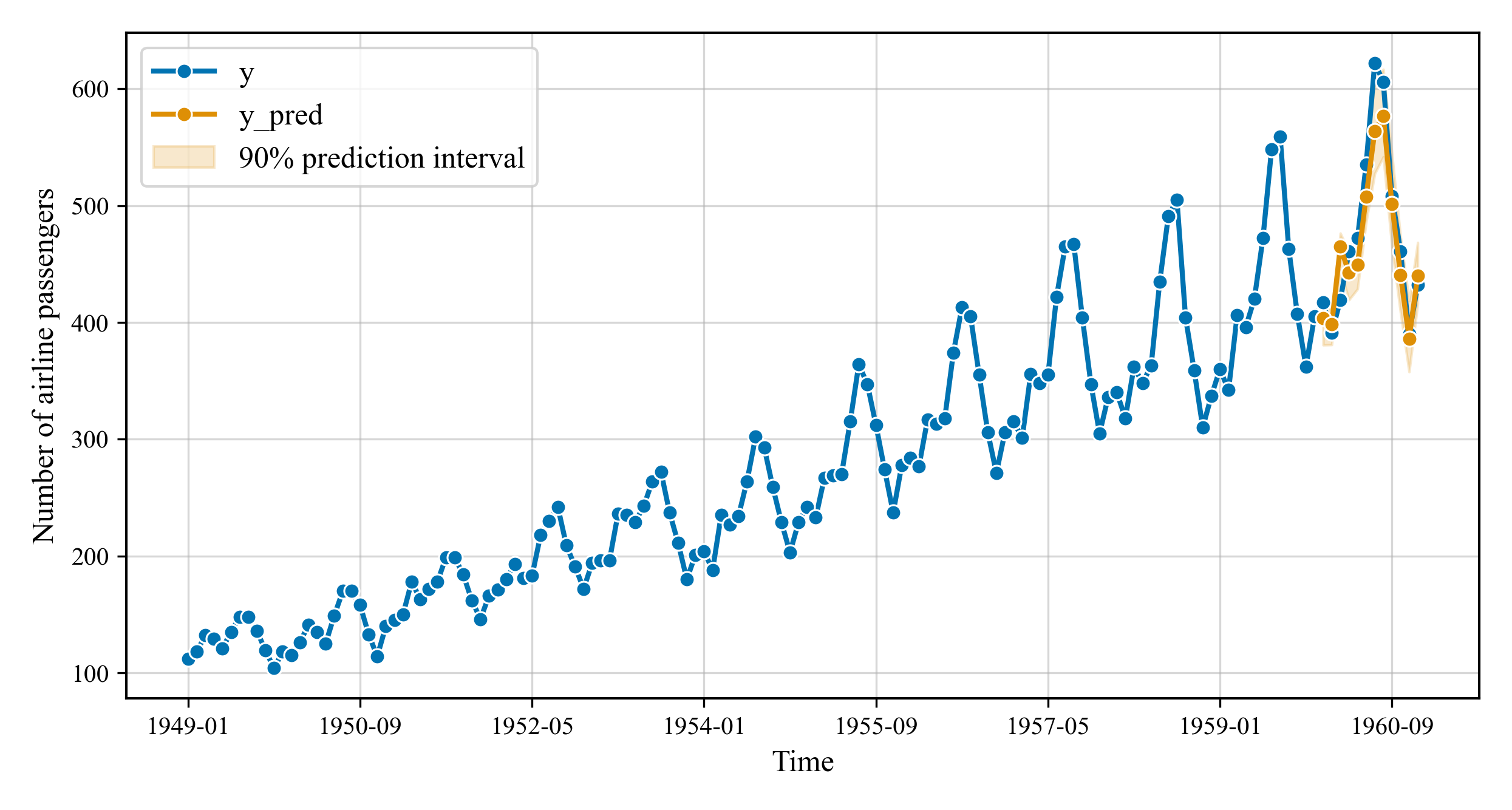}
    \caption{Exemplary usage of \texttt{tsbootstrap} and \texttt{sktime}; output from \cref{code:sktime}.}
    \label{fig:exemplary_forecast}
\end{figure}

The resulting forecast together with the intervals are shown in \Cref{fig:exemplary_forecast}.

\section{Conclusion}\label{sec:conclusion}
We have demonstrated that \texttt{tsbootstrap} advances the field of time series analysis by offering a robust and user-friendly \texttt{python} package designed for sophisticated bootstrapping techniques. Its comprehensive suite of methods not only addresses the specific needs of time series data but also enhances their analysis with a focus on usability and integration. This tool can play a crucial role in fields where accurate uncertainty quantification is essential, such as finance, meteorology, and epidemiology. Through \texttt{tsbootstrap}, researchers and practitioners can better predict and manage risks, ultimately leading to more informed decision-making.

We invite the community to participate in the ongoing development of \texttt{tsbootstrap}. We are particularly keen on collaborations that explore new domains of application or enhance our understanding of time series analysis challenges. By joining our efforts, you can help shape the future of bootstrapping techniques in time series analysis. The continuous development and enhancement of \texttt{tsbootstrap}, as highlighted in its GitHub repository, and its close integration with \texttt{sktime}, ensures its relevance and utility in the evolving landscape of time series analysis. We are excited to see how \texttt{tsbootstrap} will continue to evolve and how you, the community, will be a part of this journey.

\section{Future Work}\label{sec:future_work}
We have a clear yet evolving roadmap for future enhancements, including the integration of more sophisticated bootstrapping methods and improved performance optimizations. These are delineated in Issue \href{https://github.com/astrogilda/tsbootstrap/issues/144}{\#144} in the \texttt{tsbootstrap} repository.

\begin{enumerate}

    \item \textbf{Performance and Scaling:} Improve handling of large datasets, by integrated with established distributed backends such as Dask, Spark, and Ray. Explore ways to optimize and profile the code for improvements in both time- and space-complexity.
    
    \item \textbf{Tuning and Automation:} Implement adaptive algorithms that dynamically adjust block sizes based on autocorrelation, and integrate fractional block lengths for precise resampling \citep{block_length_selection}. Additionally, implement automated feedback mechanisms to refine the resampling process by analyzing real-time dataset characteristics and iteratively adjusting to minimize discrepancies between original and bootstrapped data.
    
    \item \textbf{Real-time and Streaming Data:} Process data in real-time by incorporating event-driven programming or reactive frameworks that handle data streams efficiently.
    
    \item \textbf{Enhanced integration with \texttt{sktime}:} Develop standardized evaluation tools that utilize \texttt{sktime}'s metrics for detailed performance comparisons between original and bootstrapped time series data. This includes establishing a repository of shared datasets for benchmarking across both libraries, and creating extensive documentation and examples that demonstrate the combined use of \texttt{tsbootstrap} and \texttt{sktime}. Additionally, introduce functionalities to generate distribution or sampler-like objects for advanced probabilistic forecasting.
    
    \item \textbf{API Extension:} Support \texttt{pandas DataFrame}s, and expand API capabilities to handle panel and hierarchical time series. At the same time, improve handling of exogenous data within bootstraps for complex models, and refine model state management to accommodate both fittable and pretrained models.

\end{enumerate}

\newpage
\vskip 0.2in
\bibliography{sample}

\end{document}